\documentclass[aps,prl,twocolumn,nofootinbib,superscriptaddress,preprintnumbers,floatfix,10pt]{revtex4-1}
\pdfoutput=1
\usepackage[bookmarks=false]{hyperref}
\usepackage[all]{hypcap}
\usepackage{graphicx}
\usepackage{amsmath}
\usepackage{xspace}
\usepackage[normalem]{ulem}

\newcommand{\refcite}[1]{Ref.~\cite{#1}}
\newcommand{\refscite}[1]{Refs.~\cite{#1}}
\newcommand{\eq}[1]{Eq.~\eqref{eq:#1}}
\newcommand{\eqs}[2]{Eqs.~\eqref{eq:#1} and \eqref{eq:#2}}

\newcommand{\fig}[1]{Fig.~\ref{fig:#1}}

\newcommand{\abs}[1]{\lvert#1\rvert}

\newcommand{\ord}[1]{\mathcal{O}(#1)}
\newcommand{\ORd}[1]{\mathcal{O}\Bigl(#1\Bigr)}

\newcommand{\df}{\mathrm{d}}
\newcommand{\img}{\mathrm{i}}

\newcommand{\qt}{\vec q_T}

\newcommand{\Ecm}{E_\mathrm{cm}}

\newcommand{\GeV}{\mathrm{GeV}}
\newcommand{\pb}{\mathrm{pb}}
\newcommand{\fb}{\mathrm{fb}}

\newcommand{\nn}{\nonumber}

\newcommand{\sing}{\mathrm{sing}}
\newcommand{\nons}{\mathrm{nons}}

\newcommand{\fo}{\mathrm{FO}}
\newcommand{\incl}{\mathrm{incl}}
\newcommand{\fid}{\mathrm{fid}}
\newcommand{\off}{\mathrm{off}}
\newcommand{\sub}{\mathrm{sub}}
\newcommand{\cut}{\mathrm{cut}}
\newcommand{\match}{\mathrm{match}}

\newcommand{\BR}{\mathcal{B}}

\newcommand{\zero}{{(0)}}
\newcommand{\one}{{(1)}}
\newcommand{\two}{{(2)}}
\newcommand{\fidpc}{\mathrm{fpc}}

\newcommand{\qT}{q_T}

\newcommand{\scetlib}{{\tt SCETlib}}
\newcommand{\nnlojet}{{\tt NNLOjet}}

\makeatletter
\g@addto@macro\bfseries{\boldmath}
\makeatother

\allowdisplaybreaks[2]

\providecommand{\papersection[1]}{\section{#1}\vspace{-1ex}}
\newcommand{\plotwidth}{\columnwidth}

\begin{document}

%%%%%%%%%%%%%%%%%%%%%%%%%%%%%%%%%%%%%%%%%%%%%%%%%%%%%%%%%%%%%%%%%%%%%%%%%%%%%%%%
% Title page
%%%%%%%%%%%%%%%%%%%%%%%%%%%%%%%%%%%%%%%%%%%%%%%%%%%%%%%%%%%%%%%%%%%%%%%%%%%%%%%%

\preprint{\vbox{\hbox{DESY 21-022}\hbox{MPP--2021--16}\hbox{MIT--CTP 5266}}}

\title{Higgs $p_T$ Spectrum and Total Cross Section with Fiducial Cuts at
Third Resummed and Fixed Order in QCD}

\author{Georgios Billis}
\affiliation{Deutsches Elektronen-Synchrotron (DESY), D-22607 Hamburg, Germany}

\author{Bahman Dehnadi}
\affiliation{Deutsches Elektronen-Synchrotron (DESY), D-22607 Hamburg, Germany}

\author{Markus A.~Ebert}
\affiliation{Max-Planck-Institut f\"ur Physik, F\"ohringer Ring 6, 80805 M\"unchen, Germany}

\author{Johannes K.~L.~Michel}
\affiliation{Center for Theoretical Physics, Massachusetts Institute of Technology, Cambridge, Massachusetts 02139, USA}

\author{Frank J.~Tackmann}
\affiliation{Deutsches Elektronen-Synchrotron (DESY), D-22607 Hamburg, Germany}

\date{February 15, 2021}

%%%%%%%%%%%%%%%%%%%%%%%%%%%%%%%%%%%%%%%%%%%%%%%%%%%%%%%%%%%%%%%%%%%%%%%%%%%%%%%%
\begin{abstract}

We present predictions for the gluon-fusion Higgs $p_T$ spectrum
at third resummed and fixed order (N$^3$LL$'+$N$^3$LO)
including fiducial cuts as required by experimental
measurements at the Large Hadron Collider.
Integrating the spectrum, we predict for the first time the total fiducial cross
section to third order (N$^3$LO) and improved by resummation. The
N$^3$LO correction is enhanced by cut-induced logarithmic effects and is not
reproduced by the inclusive N$^3$LO correction times a lower-order acceptance.
These are the highest-order predictions of their kind achieved so far at a hadron collider.

\end{abstract}
%%%%%%%%%%%%%%%%%%%%%%%%%%%%%%%%%%%%%%%%%%%%%%%%%%%%%%%%%%%%%%%%%%%%%%%%%%%%%%%%

\maketitle

%%%%%%%%%%%%%%%%%%%%%%%%%%%%%%%%%%%%%%%%%%%%%%%%%%%%%%%%%%%%%%%%%%%%%%%%%%%%%%%%
\papersection{Introduction}
%%%%%%%%%%%%%%%%%%%%%%%%%%%%%%%%%%%%%%%%%%%%%%%%%%%%%%%%%%%%%%%%%%%%%%%%%%%%%%%%

Fiducial and differential cross-section measurements of the discovered Higgs
boson~\cite{Aad:2012tfa, Chatrchyan:2012ufa} provide the most model-independent
way to study Higgs production at the Large Hadron Collider. They are thus
central to its physics program~\cite{Aad:2013wqa, Aad:2014lwa,
Aad:2014tca, Aad:2016lvc, Aaboud:2017oem, Aaboud:2018xdt, ATLAS:2020wny,
Khachatryan:2015rxa, Khachatryan:2015yvw, Khachatryan:2016vnn, Sirunyan:2018kta,
Sirunyan:2018sgc, CMS:2020dvg} and will remain so in the
future~\cite{Cepeda:2019klc}.

Theoretical predictions for the dominant gluon-fusion ($gg\to H$) Higgs
production mode suffer from large perturbative corrections. This has led to the
calculation of the total inclusive production cross section to third order
(N$^3$LO)~\cite{Dawson:1990zj, Djouadi:1991tka, Spira:1995rr, Harlander:2002wh,
Anastasiou:2002yz, Ravindran:2003um, Anastasiou:2015vya, Anastasiou:2016cez,
Mistlberger:2018etf}, which is made possible by treating the decay of the Higgs
boson fully inclusively. Unfortunately, this also makes it a primarily
theoretical quantity; one that cannot be measured in experiment. The
experimental measurements necessarily involve kinematic selection and acceptance
cuts on the Higgs decay products, which reduce the cross section by an $\ord{1}$
amount. Therefore, any comparison of theory and experiment always involves a
prediction of the \emph{fiducial} cross section, i.e., the cross section within
the experimental acceptance. Currently, the fiducial cross section for $gg\to H$
is only known to second order (NNLO). A key challenge is to calculate it at N$^3$LO, which we
do here for the first time. To be specific, we consider
$H\to \gamma\gamma$ with the fiducial cuts used by ATLAS~\cite{Aaboud:2018xdt, ATLAS:2019jst},
%%%
\begin{align} \label{eq:cuts}
p_T^{\gamma1} &\geq 0.35\,m_H
\,,\qquad
p_T^{\gamma2} \geq 0.25\,m_H
\,,\nn \\
\abs{\eta^\gamma} &\leq 1.37
\quad\text{or}\quad
1.52 \leq \abs{\eta^\gamma} \leq 2.37
\,.\end{align}
%%%

Arguably, the most important differential cross section of the Higgs boson is
its transverse-momentum ($q_T$) distribution, serving as a benchmark spectrum in
many experimental analyses. At finite $q_T$, it is known to
NNLO$_1$~\cite{deFlorian:1999zd, Ravindran:2002dc, Glosser:2002gm, Chen:2014gva,
Boughezal:2015aha, Boughezal:2015dra, Caola:2015wna, Chen:2016zka, Campbell:2019gmd,
Chen:2019wxf}, i.e., from calculating $H+1$ parton to NNLO, including
fiducial cuts, which is an important ingredient for our results. For
$q_T\ll m_H$, the $q_T$ spectrum contains large Sudakov logarithms of $q_T/m_H$,
which must be resummed to all orders in perturbation theory to obtain precise
and reliable predictions. So far, this resummation has been achieved to NNLL$'$
and N$^3$LL~\cite{Bozzi:2005wk, Becher:2012yn, Neill:2015roa, Bizon:2017rah,
Chen:2018pzu, Bizon:2018foh, Gutierrez-Reyes:2019rug, Becher:2020ugp}, which
include second and third order evolution and which capture in particular all
$\ord{\alpha_s^2}$ contributions that are singular for $q_T\to 0$. This
is also the basis of the $q_T$ subtraction method for NNLO calculations~\cite{Catani:2007vq}.

In this Letter, we obtain for the first time the resummed $q_T$ spectrum at
N$^3$LL$'+$N$^3$LO, both inclusively and with fiducial cuts. This is the
highest order achieved to date for a differential distribution at a hadron
collider. Compared to N$^3$LL, the resummation at N$^3$LL$'$ incorporates the
complete $\ord{\alpha_s^3}$ singular structure for $q_T\to 0$, i.e.,
all 3-loop virtual and corresponding real corrections, allowing us to consistently match
to N$^3$LO.
We incorporate the fiducial cuts in the resummed $q_T$ spectrum following the
recent analysis in \refcite{Ebert:2020dfc}. This allows us to also resum large,
so-called fiducial power corrections induced by the fiducial
cuts~\cite{Ebert:2019zkb, Ebert:2020dfc}, and eventually to predict the total
fiducial cross section at N$^3$LO from the integral of the resummed fiducial
$q_T$ spectrum.
This constitutes the first complete application of $q_T$ subtractions at this order.
(For earlier results and discussions see \refscite{Cieri:2018oms, Billis:2019vxg}.)

The total inclusive cross section can be considered independently
of the $q_T$ spectrum. In particular, the $\qT$ resummation effects in the
inclusive spectrum formally cancel in its integral.
This cancellation is broken by the fiducial power corrections, causing leftover
logarithmic contributions in the total fiducial cross section which worsen its
perturbative behavior.
They are resummed by integrating the resummed spectrum, restoring the
perturbative convergence. Hence, the fiducial $q_T$ spectrum
is now the more fundamental quantity, while the total fiducial cross section
becomes a derived quantity.

%%%%%%%%%%%%%%%%%%%%%%%%%%%%%%%%%%%%%%%%%%%%%%%%%%%%%%%%%%%%%%%%%%%%%%%%%%%%%%%%
\papersection{$\qT$ Resummation with Fiducial Power Corrections}
%%%%%%%%%%%%%%%%%%%%%%%%%%%%%%%%%%%%%%%%%%%%%%%%%%%%%%%%%%%%%%%%%%%%%%%%%%%%%%%%

We work in the narrow-width limit and factorize
the cross section into Higgs production and decay,
%%%
\begin{equation} \label{eq:xs_fiducial}
\frac{\df\sigma}{\df\qT}
= \int\!\df Y A(\qT, Y; \Theta)\, W(\qT, Y)
\,.\end{equation}
%%%
Since the Higgs is a scalar boson, \eq{xs_fiducial} contains a single
hadronic structure function $W(\qT, Y)$ encoding the $gg\to H$ production process.
As $W$ is a Lorentz-scalar function and inclusive over the hadronic final state,
it can only depend on the Higgs momentum $q^\mu$ and the proton momenta $P_{a,b}^\mu$
via $q^2 = m_H^2$ and $2q\cdot P_{a,b} = \Ecm \sqrt{m_H^2 + \qt^{\,2}} e^{\mp Y}$, where $Y$ and $\qt$
are the Higgs rapidity and transverse momentum.
Its only nontrivial dependence is thus on $Y$ and $\qT = \abs{\qt}$.
The function $A(q_T, Y;\Theta)$ in \eq{xs_fiducial} encodes the Higgs decay
including fiducial cuts, collectively denoted by $\Theta$.
It corresponds to the fiducial acceptance point by point in $\qT$ and $Y$,
such that in the inclusive case without cuts $A_\incl(q_T, Y) = 1$. Hence, $W$
is equivalent to the inclusive spectrum.

Let us expand the $q_T$ spectrum in powers of $q_T/m_H$,
%%%
\begin{alignat}{4} \label{eq:xs_schematic}
\frac{\df\sigma}{\df\qT}
&= \qquad\frac{\df\sigma^\zero}{\df \qT}
   &&+\quad\frac{\df\sigma^\one}{\df\qT}
   &&+\quad\frac{\df\sigma^\two}{\df\qT}
   &&+\,\,\cdots
\nn\\
&\sim \frac{1}{\qT} \biggl[\,\,\ord{1}
&&+ \ORd{\frac{\qT}{m_H}}
&&+ \ORd{\frac{\qT^2}{m_H^2}}
&&+\,\, \cdots\biggr]
\,.\end{alignat}
%%%
The singular, leading-power term $\df\sigma^\zero/\df\qT$ scales as $1/\qT$ and dominates
for $\qT\ll m_H$. It contains $\delta(\qT)$ and $[\ln^n(\qT/m_H)/\qT]_+$ distributions
encoding the cancellation of real and virtual infrared singularities at $\qT = 0$.
The $\df\sigma^{(n\geq 1)}/\df q_T$ are called power corrections.

Because of azimuthal symmetry, $W(\qT, Y)$ only receives quadratic power
corrections~\cite{Ebert:2018gsn, Ebert:2020dfc},
%%%
\begin{equation}
W(\qT, Y) = W^\zero(\qT, Y) + W^\two(\qT, Y) + \dotsb
\,,\end{equation}
%%%
where $W^\zero \sim 1/\qT$ contains the singular terms.
The acceptance corrections are finite at $\qT = 0$, but the fiducial cuts
generically break azimuthal symmetry such that it receives linear power
corrections~\cite{Ebert:2019zkb, Ebert:2020dfc},
%%%
\begin{equation}
A(\qT, Y; \Theta) = A(0, Y;\Theta)\Bigl[1 + \ORd{\frac{\qT}{m_H}}\Bigr]
\,.\end{equation}
%%%
The strict leading-power spectrum is thus given by
%%%
\begin{equation} \label{eq:LP}
\frac{\df\sigma^\zero}{\df\qT} = \int\!\df Y\, A(0, Y;\Theta)\, W^\zero(\qT, Y)
\,.\end{equation}
%%%

The fiducial power corrections,
%%%
\begin{equation} \label{eq:fidpc}
\!\,\frac{\df\sigma^\fidpc}{\df\qT} \! = \!\int\!\df Y \Bigl[A(\qT, Y;\Theta) - A(0, Y;\Theta)\Bigr] W^\zero(\qT, Y)
,\!\!\!\end{equation}
%%%
were analyzed in \refscite{Ebert:2019zkb, Ebert:2020dfc}.
They include all linear power corrections $\df\sigma^\one/\df\qT$ and are absent
in the inclusive spectrum. They are quite subtle, and can be further enhanced
to $\ord{\qT/p_L}$, where $p_L$ is an effective kinematic scale set by the fiducial cuts
with typically $p_L \ll m_H$. This prohibits expanding $A(\qT, Y;\Theta)$
even for $\qT\ll m_H$ once $\qT \sim p_L$.
For example, for the photon $p_T^\cut$, $p_L \sim m_H - 2p_T^\cut$.
For the cuts in \eq{cuts}, the expansion of $A$ starts
failing for $\qT\gtrsim 10\,\GeV$,
where the inclusive power corrections from $W^\two$ are at the few-percent level.
It is thus critical to use the exact $\qT$-dependent
acceptance and take
%%%
\begin{equation} \label{eq:sing_def}
\frac{\df\sigma^\sing}{\df\qT}
= \int\!\df Y\, A(\qT, Y;\Theta)\, W^\zero(\qT, Y)
\end{equation}
%%%
as the leading ``singular'' contribution at small $\qT$, corresponding to the sum of
\eqs{LP}{fidpc}. The remaining ``nonsingular'' contributions,
%%%
\begin{align}
\frac{\df\sigma^\nons}{\df\qT}
= \int\!\df Y\, A(\qT, Y;\Theta)\, \Bigl[W^\two(\qT, Y) + \dotsb\Bigr]
\,,\end{align}
%%%
are then suppressed by $\ord{\qT^2/m_H^2}$.

In general, and for the ATLAS cuts in particular,
$A(\qT, Y; \Theta)$ is a very nasty function given by a
boosted phase-space integral over a conjunction of complicated
$\theta$ functions encoding all cuts. Nevertheless,
using a dedicated semianalytic algorithm we are able to evaluate it
with sufficient numerical speed and accuracy.

As $A$ itself does not contain large logarithms,
resumming $W^\zero$ in \eq{sing_def} correctly resums also
the fiducial power corrections in \eq{fidpc} to the same order~\cite{Ebert:2020dfc}.
The resummation of $W^\zero$ is equivalent to that of the leading-power inclusive spectrum,
and follows from its factorization theorem originally derived
in \refscite{Collins:1981uk, Collins:1981va, Collins:1984kg} or
equivalent formulations~%
\cite{Catani:2000vq, Collins:1350496, Becher:2010tm, GarciaEchevarria:2011rb, Chiu:2012ir, Collins:2012uy, Li:2016axz}.
We employ soft-collinear effective theory~\cite{Bauer:2000ew, Bauer:2000yr, Bauer:2001ct, Bauer:2001yt, Bauer:2002nz}
with rapidity renormalization~\cite{Chiu:2011qc, Chiu:2012ir} using the exponential regulator~\cite{Li:2016axz},
where
%%%
\begin{align} \label{eq:W0_fact}
& W^\zero(\qT, Y)
\\ & \quad
= H(m_H^2, \mu) \int\!\df^2\vec k_a \, \df^2\vec k_b \, \df^2 \vec k_s\,\delta\bigl(\qT - \abs{\vec k_a + \vec k_b + \vec k_s}\bigr)
\nn\\&\qquad \times
   B_g^{\mu\nu}(x_a, \vec k_a, \mu, \nu) \,
   B_{g\, \mu\nu}(x_b, \vec k_b, \mu, \nu) \, S(\vec k_s, \mu, \nu)
\nn\,.\end{align}
%%%
The hard function $H$ contains the
effective $gg\to H$ form factor. The beam functions $B_g^{\mu\nu}$
describe collinear radiation with total transverse momentum $\vec k_{a,b}$ and
longitudinal momentum fractions $x_{a,b} = (m_H/\Ecm) e^{\pm Y}$.
The soft function $S$ describes soft
radiation with total transverse momentum $\vec k_s$.

All functions in \eq{W0_fact} are renormalized objects, with $\mu$ and $\nu$
denoting their virtuality and rapidity renormalization scales.
The all-order resummation follows by first evaluating each function
at fixed order at its own natural boundary scale(s)
$\mu_{H,B,S}$, $\nu_{B,S}$. These boundary conditions are then evolved to a
common (arbitrary) point
in $\mu$ and $\nu$ by solving the coupled system of renormalization group
equations. The exact solution for the $\qT$ distribution is formally
equivalent~\cite{Ebert:2016gcn} to the canonical solution in conjugate ($b_T$) space,
which is the approach we follow here; see \refscite{Ebert:2016gcn,
Billis:2019evv, Ebert:2020dfc} for details. At N$^3$LL$'$ (N$^3$LL) we require the
N$^3$LO (NNLO) boundary conditions for the hard~\cite{Harlander:2000mg, Baikov:2009bg,
Lee:2010cga, Gehrmann:2010ue, Ebert:2017uel}
and beam and soft
functions~\cite{Luebbert:2016itl, Li:2016ctv, Billis:2019vxg, Luo:2019bmw,
Ebert:2020yqt, Luo:2020epw}, the 3-loop noncusp anomalous
dimensions~\cite{Moch:2005tm, Idilbi:2005ni, Berger:2010xi, Luebbert:2016itl,
Li:2016ctv, Vladimirov:2016dll, Billis:2019vxg}, and the 4-loop $\beta$
function~\cite{Tarasov:1980au, Larin:1993tp, vanRitbergen:1997va, Czakon:2004bu}
and gluon cusp anomalous dimension~\cite{Korchemsky:1987wg, Vogt:2004mw,
Lee:2019zop, Henn:2019rmi, Bruser:2019auj, Henn:2019swt, vonManteuffel:2020vjv}.
At NNLL, all ingredients enter at one order lower than at N$^3$LL.

The 3-loop beam function boundary terms have been computed only
recently~\cite{Ebert:2020yqt, Luo:2020epw}. They involve a plethora of harmonic
polylogarithms up to weight five with nontrivial rational prefactors, which must
be convolved against the PDFs. This makes a naive implementation too slow and
numerically unstable. Instead, we obtain fast numerical
implementations for all kernels at close to double precision using a dedicated
algorithm that separates an entire kernel into pieces with only single branch
cuts, which then admit suitable, fast-converging logarithmic expansions around
$z = 0$ and $z = 1$.

The hard function $H$ contains timelike
logarithms $\ln[(-m_H^2 - \img 0)/\mu^2)]$, which are resummed
by using an imaginary boundary scale $\mu_H = -\img m_H$. This significantly improves
the perturbative convergence compared to the spacelike choice $\mu_H = m_H$~\cite{Parisi:1979xd,
Sterman:1986aj, Magnea:1990zb, Eynck:2003fn, Ahrens:2008qu}.
It is advantageous to apply this timelike resummation not just to $W^\zero$, which
contains $H$ naturally, but also to the full $W(\qT, Y)$,
as demonstrated for the rapidity spectrum in \refcite{Ebert:2017uel},
or equivalently the nonsingular corrections, as in similar contexts~\cite{Berger:2010xi, Stewart:2013faa}.
To do so, we take~\cite{Ebert:2017uel}
%%%
\begin{equation} \label{eq:W_rest}
W(\qT, Y)
= H(m_H^2, \mu_\fo)\, \biggl[\frac{W(\qT, Y)}{H(m_H^2, \mu_\fo)}\biggr]_\fo
\,,\end{equation}
%%%
and analogously for $\df\sigma^\nons/\df\qT$.
The ratio in square brackets is expanded to fixed order in $\alpha_s(\mu_\fo)$,
while $H(m_H^2, \mu_\fo)$ in front is evolved from $\mu_H$ to $\mu_\fo$ at
the same order as in \eq{W0_fact}.
This yields substantial improvements up to $\qT\sim 200\,\GeV$,
which is not unexpected, as $W^\two$ will contain $H$ in parts of its factorization.
(Beyond $\qT\gtrsim 200\,\GeV$, a dynamic hard scale $\sim \qT$ becomes more
appropriate and the heavy-top limit breaks down, indicating that the
hard interaction has become completely unrelated to the $H+0$-parton process.)

The fixed-order coefficients of $\df\sigma^\nons/\df\qT$ for $\qT > 0$ are obtained as
%%%
\begin{align} \label{eq:nons_fo}
\frac{\df\sigma^\nons_\fo}{\df\qT}
= \frac{\df\sigma_{\fo_1}}{\df\qT} - \frac{\df\sigma^\sing_\fo}{\df\qT}
\,.\end{align}
%%%
At N$^n$LO ($\equiv$ N$^n$LO$_0$), or $\ord{\alpha_s^n}$ relative to the LO Born
cross section, we need the full spectrum at N$^{n-1}$LO$_1$.
At LO$_1$ and NLO$_1$,
we integrate our own analytic implementation of $W(\qT, Y)$ against $A(\qT, Y;\Theta)$,
allowing us to reach $10^{-4}$ relative
precision down to $\qT = 0.1\,\GeV$ at little computational cost. At
NLO$_1$, we implement results from \refcite{Dulat:2017brz} after performing
the necessary renormalization. The implementation is checked against the
numerical code from \refcite{Glosser:2002gm}.
At NNLO$_1$, we use existing results~\cite{Chen:2018pzu, Bizon:2018foh} from
\nnlojet~\cite{Chen:2014gva, Chen:2016zka} (see below).

The final resummed $\qT$ spectrum is then given by
%%%
\begin{align} \label{eq:matching}
\frac{\df\sigma}{\df\qT}
&= \frac{\df\sigma^\sing}{\df\qT} + \frac{\df\sigma^\nons}{\df\qT}
\,.\end{align}
%%%
While for $\qT \ll m_H$, the singular and nonsingular contributions can be considered
separately, this separation becomes meaningless for $\qT \sim m_H$. To obtain a valid prediction
there, the $\qT$ resummation is switched off, only keeping the timelike resummation,
by choosing common boundary scales $\mu_{S,B} = \nu_{S,B} = \img \mu_H = \mu_\fo$, such
that singular and nonsingular exactly recombine at fixed order into the full result.
We use $\qT$-dependent profile scales~\cite{Stewart:2013faa, Lustermans:2019plv, Ebert:2020dfc}
to enforce the correct $\qT$ resummation for $\qT\ll m_H$ and smoothly turn it off
toward $\qT\sim m_H$.

%-------------------------------------------------------------------------------
\begin{figure}[t!]
\includegraphics[width=\plotwidth]{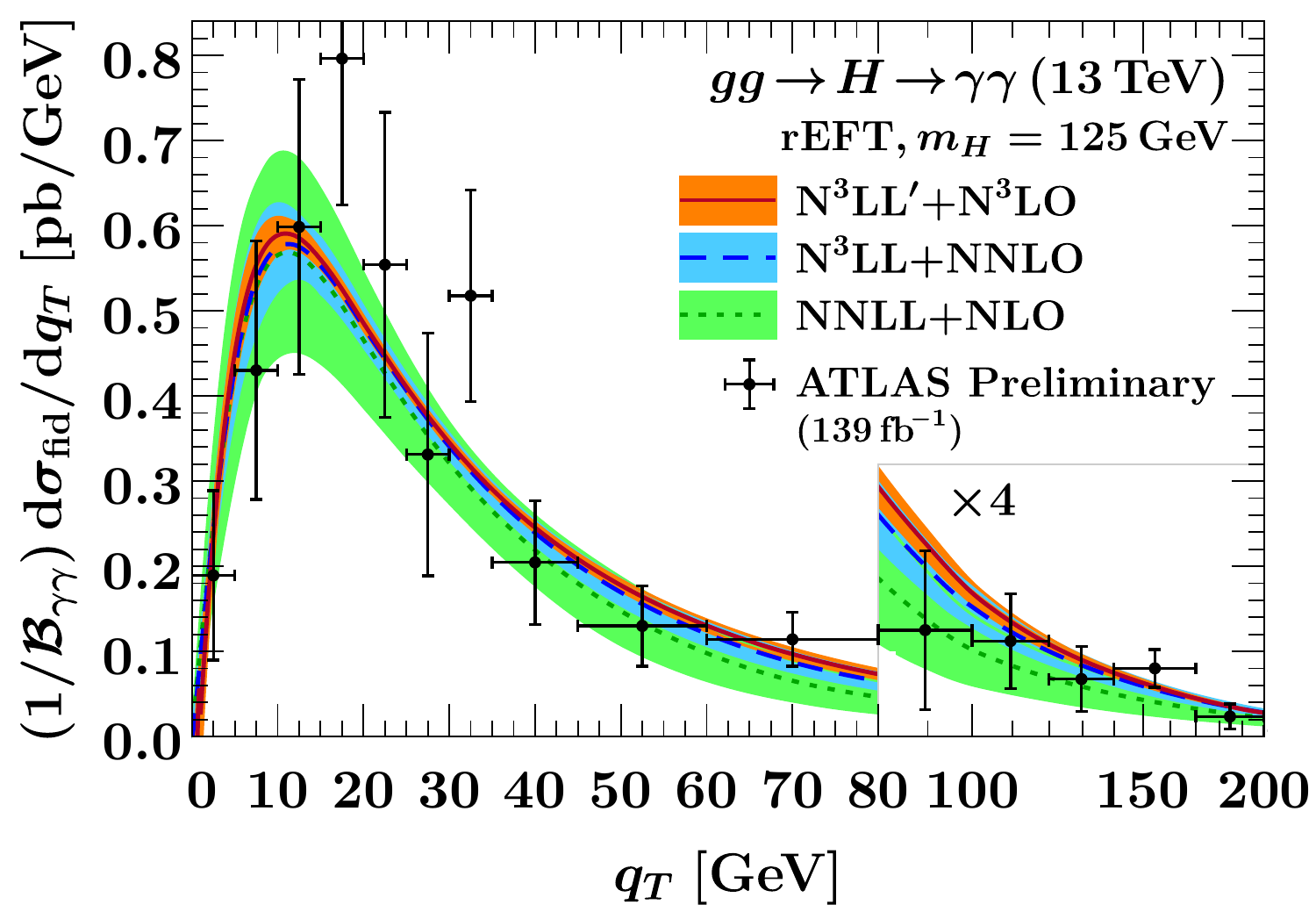}%
\\%
\includegraphics[width=\plotwidth]{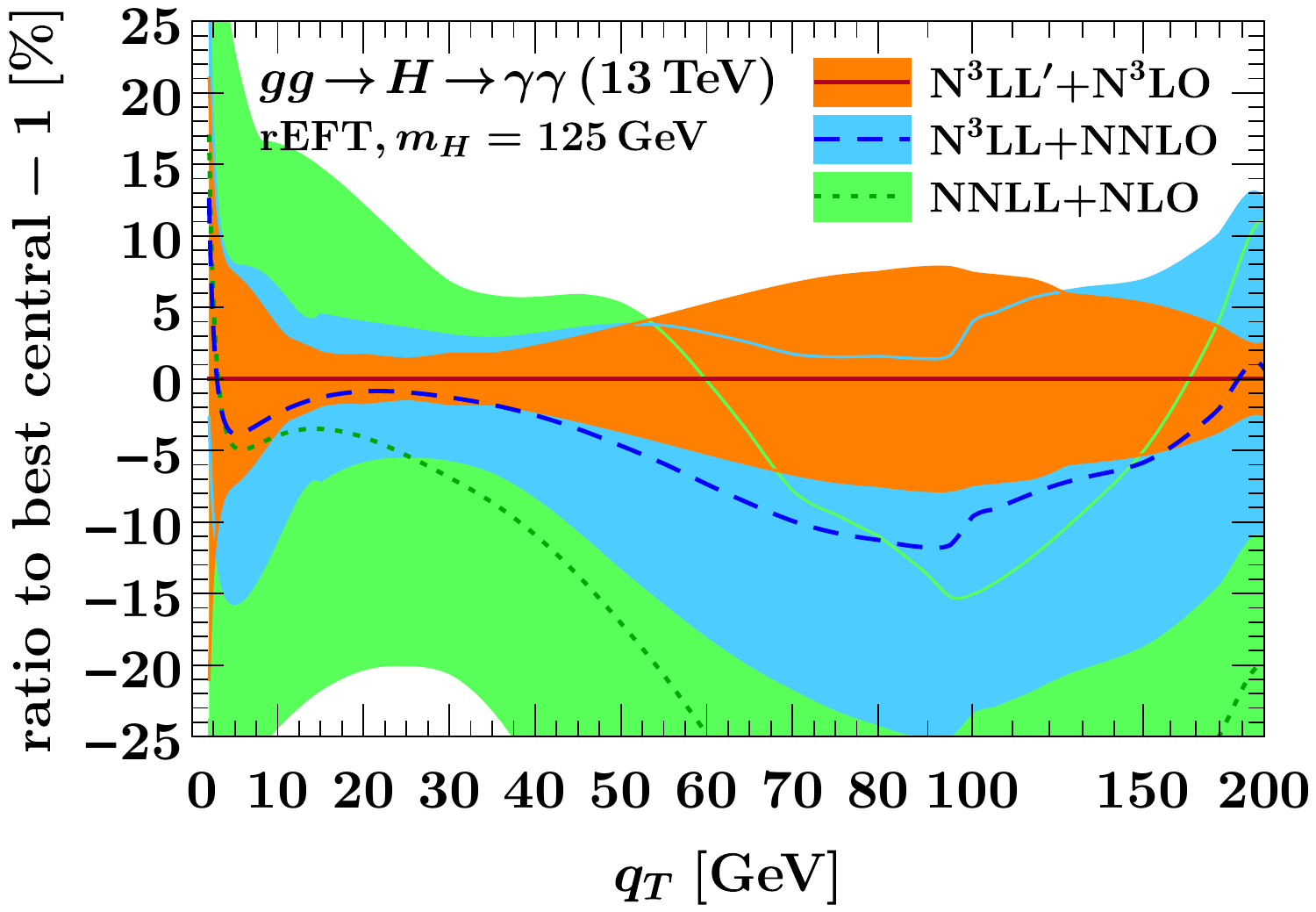}%
\vspace{-1.5ex}
\caption{The $gg\to H$ $q_T$ spectrum up to N$^3$LL$'+$N$^3$LO compared to
preliminary ATLAS measurements~\cite{ATLAS:2019jst}.
}
\vspace{-1.5ex}
\label{fig:spectrum}
\end{figure}
%-------------------------------------------------------------------------------

We identify several sources of perturbative uncertainties, namely
fixed-order ($\Delta_\fo$),
$\qT$ resummation ($\Delta_{\qT}$),
timelike resummation ($\Delta_\varphi$),
and matching uncertainties ($\Delta_\match$),
which are estimated via appropriate scale variations as detailed in
\refscite{Ebert:2020dfc, Ebert:2017uel}.
They are considered independent sources and are consequently added
in quadrature to obtain the total uncertainty.
We neglect nonperturbative effects at small $\qT$, which
are expected to be $\sim\Lambda_{\rm QCD}^2/\qT^2$ and smaller
than the current perturbative uncertainties.

Our numerical results are obtained with \scetlib~\cite{scetlib}.
We use the PDF4LHC15 NNLO parton distribution functions (PDFs)~\cite{Butterworth:2015oua},
$\alpha_s(m_Z) = 0.118$,
$\mu_\fo = m_H = 125\,\GeV$.
We work in the heavy-top limit rescaled with the exact LO dependence on $m_t = 172.5\,\GeV$
(rEFT).
By default, we exclude the $H\to\gamma\gamma$ branching ratio ($\BR_{\gamma\gamma}$)
from our predictions, $\sigma \equiv \sigma_\fid/\BR_{\gamma\gamma}$.
Our $\qT$ spectrum at N$^3$LL$'+$N$^3$LO is presented in \fig{spectrum}, showing
excellent perturbative convergence. Below $\qT\lesssim 10\,\GeV$,
this would not be the case without resumming the fiducial power corrections.
We also compare to preliminary ATLAS measurements~\cite{ATLAS:2019jst}, for which
we subtract the non-gluon-fusion background and divide by the photon isolation
efficiency~\cite{Aaboud:2018xdt} and $\BR_{\gamma\gamma}$.

%%%%%%%%%%%%%%%%%%%%%%%%%%%%%%%%%%%%%%%%%%%%%%%%%%%%%%%%%%%%%%%%%%%%%%%%%%%%%%%%
\papersection{Total Fiducial Cross Section}
%%%%%%%%%%%%%%%%%%%%%%%%%%%%%%%%%%%%%%%%%%%%%%%%%%%%%%%%%%%%%%%%%%%%%%%%%%%%%%%%

If (and only if) the singular distributional structure of $\df\sigma^\zero/\df\qT$ is known,
the $\qT$ spectrum can be integrated to obtain the total cross section. This is
the basis of $\qT$ subtractions~\cite{Catani:2007vq},
%%%
\begin{equation} \label{eq:subtractions}
\sigma
= \sigma^\sub(\qT^\off) + \int\!\df\qT
\biggl[\frac{\df\sigma}{\df\qT} - \frac{\df\sigma^\sub}{\df\qT}\theta(\qT \leq \qT^\off)\biggr]
.\end{equation}
%%%
Here, $\df\sigma^\sub = \df\sigma^\zero[1 + \ord{\qT/m_H}]$ contains the singular
terms, with $\sigma^\sub(\qT^\off)$ its distributional integral over $\qT \leq \qT^\off$,
while the term in brackets is numerically integrable.
Taking $\sigma^\sub \equiv \sigma^\sing$, we get
%%%
\begin{equation} \label{eq:qTintegral}
\sigma
= \sigma^\sing(\qT^\off) + \int_0^{\qT^\off}\!\!\df\qT\, \frac{\df\sigma^\nons}{\df\qT}
+ \int_{\qT^\off}\!\df\qT\, \frac{\df\sigma}{\df\qT}
\,,\end{equation}
%%%
which is exactly the integral of \eq{matching}.
The subtractions here are differential in $\qT$, where $\qT^\off \sim 10-100\,\GeV$
determines the range over which they act and exactly cancels between all terms.

To integrate $\df\sigma^\nons/\df\qT$ in \eq{qTintegral} down to $\qT = 0$,
we parametrize the fixed-order coefficients in \eq{nons_fo} by their leading behavior,
%%%
\begin{equation} \label{eq:fit}
\qT\frac{\df\sigma^\nons_\fo}{\df\qT}\bigg\vert_{\alpha_s^n}
\!\!= \frac{\qT^2}{m_H^2} \!\sum_{k = 0}^{2n-1} \! \Bigl(a_k + b_k \frac{\qT}{m_H} + \dotsb\Bigr) \ln^k\!\frac{\qT^2}{m_H^2}
,\end{equation}
%%%
and perform a fit to this parametrization, which we then integrate analytically.
We follow the fit procedure discussed in detail in \refscite{Moult:2016fqy, Moult:2017jsg}
including the selection of fit parameters and range, extended to the present case.
In particular, to obtain reliable, unbiased fit results, we must account for the uncertainties
in the parametrization from yet higher-power corrections. This is done by including
the next higher-power coefficients ($b_k, \ldots$) as nuisance parameters
to the extent required by the fit range and precision. In the fiducial case, all
$b_k$ coefficients are required. The fit has been validated extensively.
As a benchmark, we correctly reproduce the $\alpha_s$ ($\alpha_s^2$)
coefficients of the total inclusive cross section to better than $10^{-5}$ ($10^{-4}$)
relative precision.

At N$^3$LO, we use existing \nnlojet\ results~\cite{Chen:2018pzu, Bizon:2018foh}
to get nonsingular data for $0.74 \,\GeV \,(4\, \GeV) \leq q_T \leq q_T^\off$
for inclusive log bins (for inclusive and fiducial linear bins).
While these data are not yet precise enough toward small $\qT$ to give a stable
fit on their own, we exploit that in the inclusive case, the known $\alpha_s^3$ coefficient of the
total inclusive cross section~\cite{Mistlberger:2018etf, Bonvini:2016frm}
provides a sufficiently strong additional constraint to obtain a reliable fit.
In the fiducial case, we exploit that the inclusive and fiducial
$a_k$ are related, arising from the same $Y$-dependent
coefficient functions integrated either inclusively or against $A(0, Y;\Theta)$.
At NLO and NNLO, their ratios lie between 0.4 and 0.55. At N$^3$LO, we thus
perform a simultaneous fit to inclusive and fiducial data, using 12 fiducial
and 8 inclusive parameters, with a loose $1\sigma$ constraint on the
fiducial $a_k$ to be 0.4--0.55 times their inclusive counterparts.
(The $b_k, \ldots$ parameters are unrelated and unconstrained.)
We stress that this does not amount to rescaling any part of the fiducial NNLO cross
section with an inclusive N$^3$LO $K$ factor. It merely tells the fit to only
consider $a_k$ of roughly the right expected size. This is sufficient to break
degeneracies and yields a stable fit, with an acceptable $\sim 0.1\,\pb$
uncertainty for the fiducial nonsingular integral ($\Delta_\nons$).

%-------------------------------------------------------------------------------
\begin{figure}[t!]
\includegraphics[width=\plotwidth]{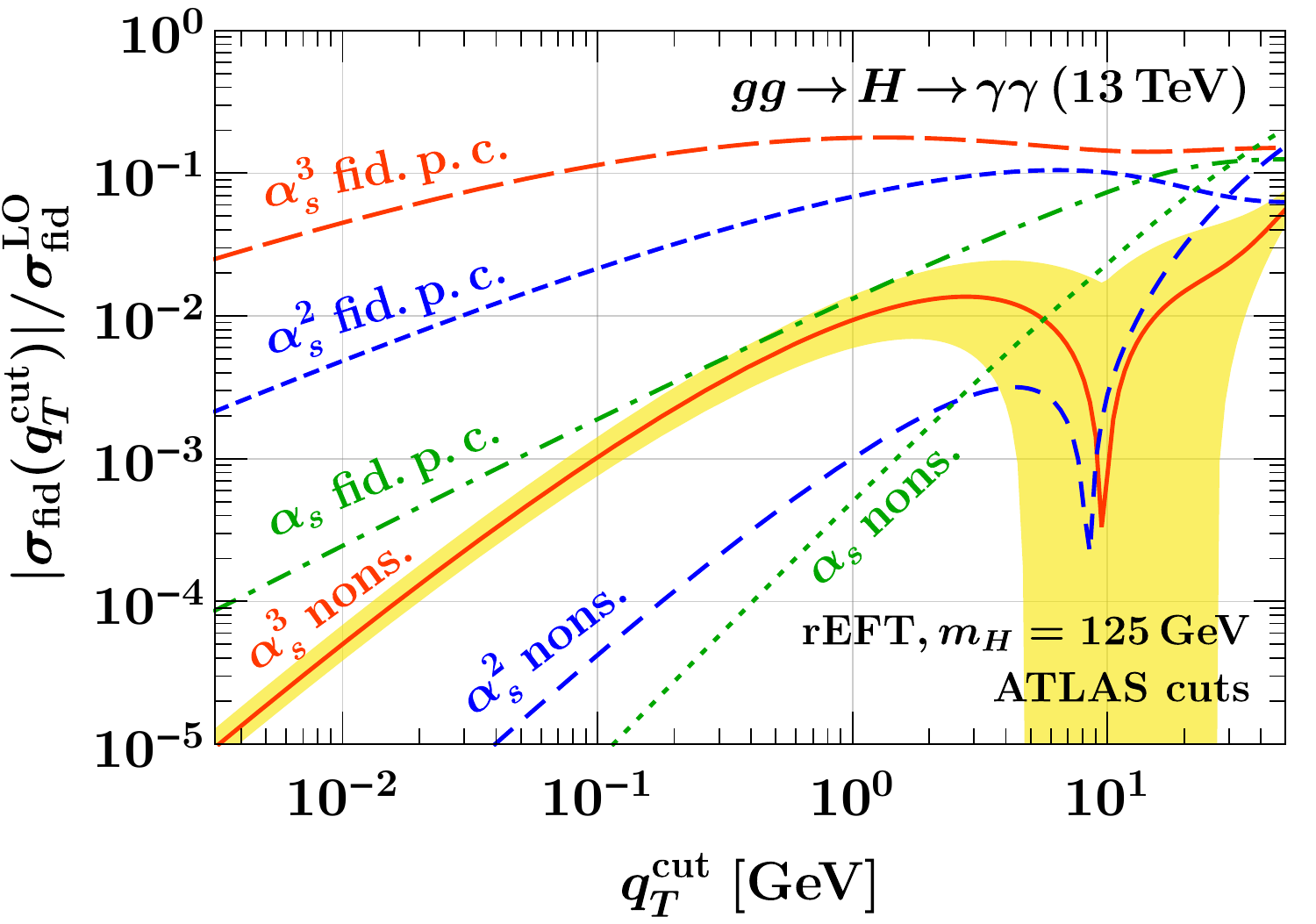}%
\vspace{-1.5ex}
\caption{Fiducial and nonsingular power corrections integrated up to $\qT \leq \qT^\cut$.
The yellow band shows $\Delta_\nons$ from the fit.}
\label{fig:nonscumulant}
\end{figure}
%-------------------------------------------------------------------------------

The often-used $\qT$ slicing approach amounts to taking $\qT^\off \to \qT^\cut \sim 1\GeV$
and simply dropping the power corrections below $\qT^\cut$.
The nonsingular and fiducial power corrections are shown in \fig{nonscumulant}.
The latter are huge at $\alpha_s^3$, and even at $\alpha_s^2$ only
become really negligible below $\qT^\cut \lesssim 10^{-2}\,\GeV$. This is why it is critical
for us to include them in the subtractions (and to resum them).
The remaining nonsingular corrections at $\alpha_s^3$ are about
10 times larger than at $\alpha_s^2$, and at $\qT^\cut = 1$--$5\,\GeV$ still
contribute $5\%$--$10\%$ of the total $\alpha_s^3$ coefficient. Together with the
current precision of the nonsingular data, this makes the above differential subtraction
procedure essential to our results.

%-------------------------------------------------------------------------------
\begin{figure}[t!]
\includegraphics[width=\plotwidth]{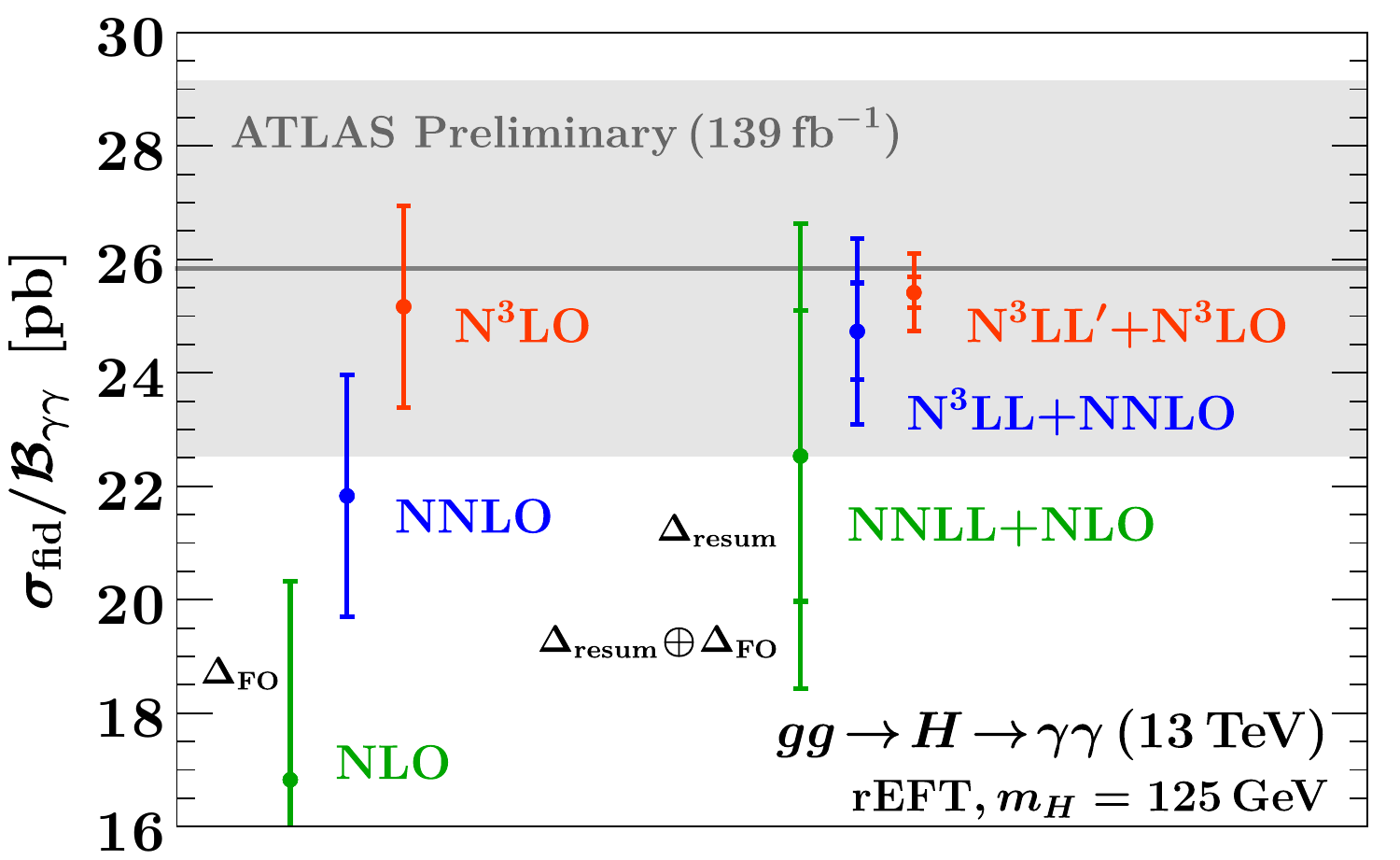}%
% \vspace{-0.5ex}
\caption{Total fiducial $gg\to H \to\gamma\gamma$ cross section
at fixed N$^3$LO (this work) and including resummation (also this work), where
$\Delta_{\rm resum} \equiv \Delta_{\qT}\oplus\Delta_\varphi\oplus\Delta_\match$,
compared to preliminary ATLAS measurements~\cite{ATLAS:2019jst}. }
\label{fig:totalxsfid}
\end{figure}
%-------------------------------------------------------------------------------

Evaluating \eq{qTintegral} either at fixed order or including
resummation, we obtain our final results for the total fiducial cross section
presented in \fig{totalxsfid}. The poor convergence at fixed order
is largely due to the fiducial power corrections. To see this,
%%%
\begin{alignat}{9} \label{eq:totalxsec_fo}
\sigma_\incl^\fo &= 13.80\,[1 &&+ 1.291 + 0.783 + 0.299]\,\pb
\,, \nn \\
\sigma^\fo_\fid/\BR_{\gamma\gamma}
&= 6.928\,[1 &&+ (1.300+0.129_\fidpc)
\nn \\ & \quad
&&+ (0.784-0.061_\fidpc)
\nn \\ & \quad
&&+ (0.331+0.150_\fidpc)] \,\pb
\,.\end{alignat}
%%%
The successive terms are the contributions from each order
in $\alpha_s$. The numbers with ``fpc'' subscript are the contributions of
the fiducial power corrections in \eq{fidpc} integrated over $q_T \leq 130 \,\GeV$.
The corrections without them are almost identical to the inclusive case.
The fiducial power corrections break this would-be universal acceptance effect,
causing a 10\% correction at NLO and NNLO and a 50\% correction at N$^3$LO and
showing no perturbative convergence.

Integrating $W^\zero$ over $\qT$, all $\qT$ logarithms and
resummation effects formally have to cancel.
(Numerically, this strongly depends on the specific implementation of resummation
and matching. We have verified explicitly that it is well satisfied in our approach.)
For the fiducial power corrections,
the nontrivial $\qT$ dependence of the acceptance spoils this cancellation and
induces residual logarithmic dependence on $p_L/m_H$ in the integral. This causes
the large corrections in \eq{totalxsec_fo}, which get resummed using the resummed
$\sigma^\sing$ in \eq{qTintegral}. Together with timelike resummation, this leads
to the excellent convergence of the resummed results in \fig{totalxsfid},
very similar to the inclusive case~\cite{Ebert:2017uel},
%%%
\begin{align}
\sigma_\incl
&= 24.16\, [1 + 0.756 + 0.207 + 0.024]\,\pb
\,,\nn \\
\sigma_\fid/\BR_{\gamma\gamma}
&= 12.89\,[1 + 0.749 + 0.171 + 0.053]\,\pb
\,.\end{align}
%%%

To conclude, our best result for the fiducial Higgs cross
section at N$^3$LL$'+$N$^3$LO for the cuts in \eq{cuts} reads
%%%
\begin{alignat}{9}
\sigma_\fid/\BR_{\gamma\gamma}
&= (25.41 &&\pm 0.59_\fo \pm 0.21_{\qT} \pm 0.17_\varphi
\nn \\ &
&&\pm 0.06_\match \pm 0.20_\nons)\,\pb
\nn \\
&= (25.41 &&\pm 0.68_\mathrm{pert})\,\pb
\,.\end{alignat}
%%%
Multiplying by $\BR_{\gamma\gamma} = (2.270 \pm 0.047)\times 10^{-3}$~\cite{Djouadi:1997yw, Bredenstein:2006rh, deFlorian:2016spz},
%%%
\begin{alignat}{9}
\sigma_\fid
&= 57.69 \, (1 &&\pm 2.7\%_\mathrm{pert} \pm 2.1\%_\BR
\\ \nn &
&&\pm 3.2\%_{\mathrm{PDF}+\alpha_s} \pm 2\%_\mathrm{EW} \pm 2\%_{t,b,c})\,\fb
\,,\end{alignat}
%%%
where we also included approximations of additional uncertainties.
The PDF$+\alpha_s$ uncertainty is taken from the
inclusive case~\cite{Anastasiou:2016cez, deFlorian:2016spz}.
For the inclusive cross section, NLO electroweak effects give a $+5\%$ correction~\cite{Actis:2008ug},
while the net effect of finite top-mass, bottom, and charm contributions is $-5\%$
(in the pole scheme we use). We can expect roughly similar acceptance corrections for both,
and therefore keep the central result unchanged but include a conservative
2\% uncertainty (40\% of the expected correction) for each effect.
Their proper treatment requires incorporating them into the resummation
framework, which we leave for future work.

%%%%%%%%%%%%%%%%%%%%%%%%%%%%%%%%%%%%%%%%%%%%%%%%%%%%%%%%%%%%%%%%%%%%%%%%%%%%%%%%
\begin{acknowledgments}
\paragraph{Acknowledgments.}
We are grateful to Xuan Chen for providing us with the \nnlojet\ results and for
communication about them. We would also like to thank our ATLAS colleagues for their
efforts in making the preliminary results of \refcite{ATLAS:2019jst} publicly available.
This work was supported in part by the Office of Nuclear Physics of the U.S.\
Department of Energy under Contract No.\ DE-SC0011090
and within the framework of the TMD Topical Collaboration,
the Deutsche Forschungsgemeinschaft (DFG) under Germany's Excellence
Strategy -- EXC 2121 ``Quantum Universe'' -- 390833306,
and the PIER Hamburg Seed Project PHM-2019-01.
\end{acknowledgments}

\paragraph{Note added.}
While finalizing this work, we became aware of complementary work computing
fiducial rapidity spectra in Higgs production at N$^3$LO using
the Projection-to-Born approach~\cite{Chen:2021isd}. The perturbative
instabilities observed there are avoided here by resumming the responsible
fiducial power corrections.

\bibliography{../paper/refs}

\end{document}